\documentclass[aps,prl,twocolumn,10pt,showpacs,superscriptaddress,amsmath,amssymb]{revtex4-1}
\usepackage{graphicx} 
\usepackage{dcolumn} 
\usepackage{bm} 
\usepackage{nicefrac} 
\usepackage{xspace} 
\usepackage{caption}
\usepackage{subcaption}
\usepackage{epstopdf}
\usepackage{color}
\usepackage{mathtools}
\begin{document}
\title{ Beyond the spin model approximation for Ramsey spectroscopy }
\date{\today}
\author{A. P. Koller}
\affiliation{JILA, NIST, and Department of Physics, University of Colorado
Boulder, CO 80309}
\author{M. Beverland}
\affiliation{Institute for Quantum Information and Matter, California Institute of Technology, MC 305-16, Pasadena CA 91125}
\author{A. V. Gorshkov}
\affiliation{Joint Quantum Institute, NIST, and University of Maryland, College Park, Maryland 20742}
\author{A. M. Rey}
\affiliation{JILA, NIST, and Department of Physics, University of Colorado
Boulder, CO 80309}
\begin{abstract}
Ramsey spectroscopy has become a powerful technique for probing non-equilibrium dynamics of internal  (pseudospin) degrees of freedom of interacting systems. In many theoretical treatments, the key to understanding the dynamics has been to assume the external (motional) degrees of freedom are decoupled from the pseudospin degrees of freedom. Determining the  validity of this approximation -- known as the spin model approximation -- has not been addressed in detail.  Here  we shed light in this direction by calculating Ramsey dynamics exactly for two interacting spin-1/2 particles in a harmonic trap. We focus on $s$-wave-interacting fermions in quasi-one and two-dimensional geometries. We find that in 1D the spin model assumption works well over a wide range of experimentally-relevant conditions, but can fail at time scales longer than those set by the mean interaction energy. Surprisingly, in 2D a modified version of the spin model is exact to first order in the interaction strength. This analysis is important for a correct interpretation of Ramsey spectroscopy and has broad applications  ranging from precision measurements to quantum information and  to  fundamental probes of many-body systems.
\end{abstract}

\pacs{03.75.Ss, 06.30.Ft, 06.20.fb, 32.30-r, 34.20.Cf}
\maketitle

\begin{figure} 
\includegraphics[scale=.3]{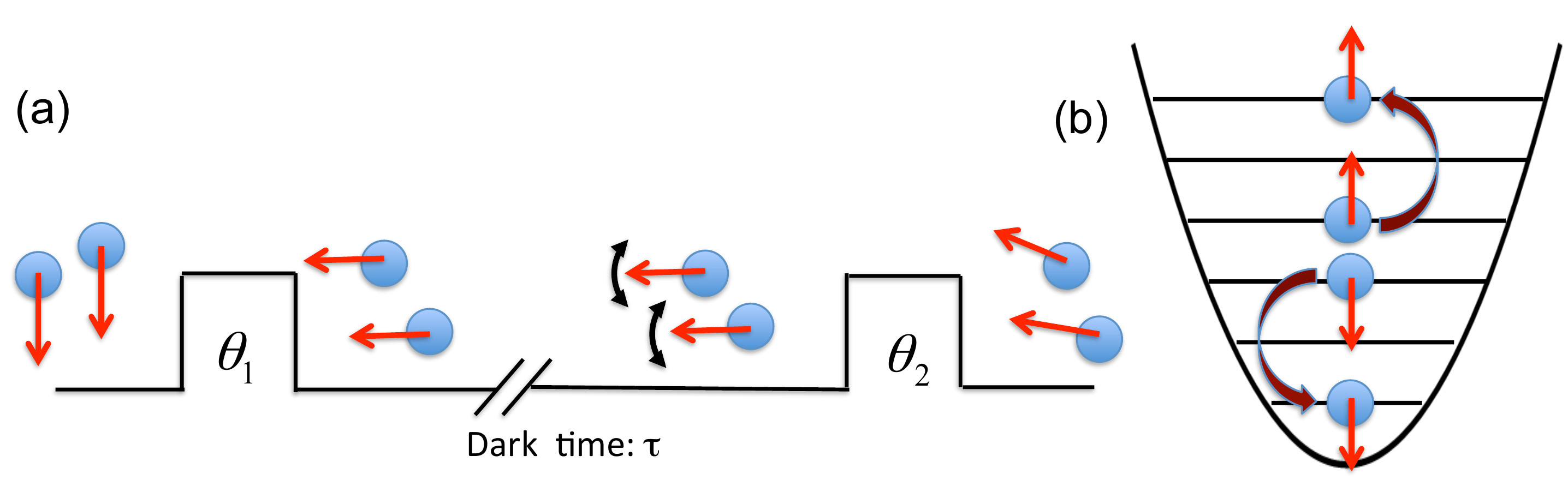}
\caption
{ \label{spinmotion}
\raggedright
(a) Ramsey spectroscopy of two interacting spin-1/2 particles.  (b) In a harmonic trap the spectrum degeneracy allows near-resonant mode-changing collisions coupled to the spin dynamics.
}
\end{figure}

Ramsey spectroscopy, a technique initially designed to interrogate microwave atomic clocks, has become an important modern tool for probing dynamics of interacting many-body systems with internal (pseudospin) degrees of freedom. Ramsey spectroscopy applies (see Fig.~\ref{spinmotion} (a)) two
strong resonant pulses to a system initially prepared in a well-defined pseudospin state, separated by a dark time of free evolution. The first  pulse initializes the pseudospin dynamics by preparing the system in a nontrivial superposition of eigenstates, i.e.\ it introduces a quantum quench \cite{Polkovnikov2011}. The second pulse reads the coherence or correlations developed during the dark time. Recently, Ramsey spectroscopy has been proposed
for extracting real-space and time correlations \cite{Knap2013,Kitagawa2010,Knap2012b,Widera2004,Kuklov2004}, characterizing  topological order \cite{Atala2012,Abanin2013}, measuring spin diffusion dynamics in bosonic  \cite{Pechkis2013,McGuirk2002,Harber2002,Lewandowski2002,Deutsch2010,Maineult2012} and fermionic systems \cite{ Du2008,Piechon2009,Natu2009,Koschorreck2013}, and as a means to probe many-body interactions in atomic, molecular, and trapped ion systems \cite{Zwierlein2003,Gupta2003,Gibble2009,Rey2009,Yu2010,Britton2012,Hazlett2013,Martin2013,Yan2013}.

Generally speaking, Ramsey spectroscopy measures the collective pseudospin and  traces out  other external degrees of freedom involved during the free evolution. In most atomic setups the latter are associated with motional degrees of freedom in the  harmonic trapping potential and/or lattice potential confining the atoms.
 The external degrees of freedom can affect  the spin dynamics in a non-trivial way, however. A great simplification could be gained if it were possible to decouple the motional and spin degrees of freedom, and reduce the many-body  dynamics down to those extracted from a pure interacting spin model. Evidence that this scenario is possible, even far from quantum degeneracy, has  been reported in  recent experiments \cite{Swallows2011,Deutsch2010,Maineult2012,Martin2013,Hazlett2013,Pechkis2013,Yan2013}, where  the observed spin dynamics corresponded to those of a pure spin Hamiltonian. These observations are opening a path for the investigation of quantum magnetism in atomic systems without the need for ultra-low temperatures.
 It is thus important to determine the parameter regime in which a pure interacting-spins picture is valid.

In this Letter we provide insight on the  validity of a pure spin model description of Ramsey spectroscopy by performing exact calculations for fermions with $s$-wave interactions and an internal pseudospin-1/2 degree of freedom, confined in quasi-1D and quasi-2D harmonic traps.
We show that the large degeneracy of the harmonic oscillator spectrum can limit the validity of the spin model to time scales less than the inverse interaction strength, due to resonant collisionally-induced excitation of spatial modes (see Fig.~\ref{spinmotion}~(b)). Cold atom experiments are protected from this problem if the  temperature is high enough that atoms  probe the actual Gaussian shape of the potential which  breaks the harmonic spectrum degeneracy. This was shown to be the case for example  in Refs.~\cite{Martin2013,Pechkis2013,Hazlett2013} where a pure spin model well described the experimental observations. At very low temperatures, Pauli blocking can also prevent mode changing collisions, as recently observed in Ref.~\cite{Krauser2013}. However, the degeneracy is a concern for intermediate temperatures at which the set of populated levels are effectively harmonic. Here we show that surprisingly, in two dimensions and to first order in the interaction strength, the full two-particle dynamics can be described in terms of an effective spin model with appropriate parameters. Our two-body calculations are not only a first step towards understanding the interplay between spin and particle motion in generic many-body ensembles, but are also  directly applicable to optical clocks that interrogate an array of 1D tube-shaped traps, each with fewer than three atoms \cite{Swallows2011,Lemke2011,Ludlow2013}.

\textit{Physical situation.}---Consider two fermions with internal degrees of freedom $\{\uparrow ,\downarrow \}$ corresponding, for instance, to the ${}^1S_0$-${}^3P_0$ electronic levels in alkaline-earth-based optical lattice clocks, and assume their interactions are primarily described by an  $s$-wave  pseudo-potential. The atoms are also illuminated by a laser beam detuned by  $\delta=\omega_L-\omega_0$ from the atomic transition $\omega_0$, with wavevector $\vec{k}$  and bare Rabi frequency $\Omega$. The two-particle Hamiltonian is then given by ${\hat H}(\vec{x}_1,\vec{x}_2)=\sum_{i=1,2}\hat{H}_{\rm L}(\vec{x}_i)+\hat{H}_{\rm{D}}(\vec{x}_1,\vec{x}_2)$: \begin{eqnarray} \label{originalH}
&&\hat{H}_{\rm L}(\vec{x}_i)=-\frac{\hbar \Omega}{2} e^{-i(\omega_L t-{\vec k} \cdot {\vec{x}_i})}  \hat{\sigma}^+_i+{\rm H.c.}\\ \nonumber
&&\hat{H}_{\rm{D}}(\vec{x}_1,\vec{x}_2)=H_{sp}(\vec{x}_1)+H_{sp}(\vec{x}_2)+g \hat{P}_{s}\delta\left(\vec{r}\right)\frac{\partial}{\partial r} r .
\end{eqnarray}
Here $\hat{H}_{\rm L}(\vec{x}_i)$ describes  the atom-laser interaction:  $\hat{\sigma}^+_i$ is the spin raising operator acting on atom $i$, and H.c.~is the Hermitian conjugate. $H_{sp}(\vec{x}_i)=-\hbar^2/(2M)\nabla_i^2+ V(\vec{x}_i)+ (\hbar\omega_0/2) \hat{\sigma}^z_i  $ is the single particle Hamiltonian with an external potential, $V$, assumed for simplicity to be independent of the internal state and separable. $H_{sp}(\vec{x}_i)$ has
 eigenfunctions $\phi_{\bf{n}}(\vec{x}_i)$ and eigenenergies $E_{\bf{n}}$ with   ${\bf n}=\{n_x,n_y,n_z\}$.
  $M$ is the particle's mass and $\hat{\sigma}^z$ the Pauli matrix. $\vec{r}=\vec{x}_1-\vec{x}_2$  is the relative coordinate,  $g=4\pi\hbar^2a_s^{\uparrow\downarrow}/M$ and   $a_s^{\uparrow\downarrow}$ the 3D $s$-wave scattering length. $\hat{P}_{s}=|s\rangle\langle s|$ is the projector into the singlet state, $|s\rangle =\frac{1}{\sqrt{2}}\left(|\!\uparrow\downarrow\rangle - |\!\downarrow\uparrow\rangle\right)$.
Only  fermions in the singlet state can interact, while spin triplet states,  $|t_{\downarrow\downarrow}\rangle = |\!\downarrow\downarrow\rangle,
|t_{\downarrow\uparrow}\rangle=\frac{1}{\sqrt{2}}\left(|\!\uparrow\downarrow\rangle+ |\!\downarrow\uparrow\rangle\right)$, and $|t_{\uparrow\uparrow}\rangle=|\!\uparrow\uparrow\rangle$ cannot experience  $s$-wave interactions.

\textit{ The spin model.}---The assumptions of the spin model are: if  there are no degeneracies in the two-atom non-interacting spectrum, i.e.~$(E_{\bf m} + E_{\bf n}) = (E_{\bf m'} + E_{\bf n'})$ occurs only for $({\bf m},{\bf n}) = ({\bf m'},{\bf n'})$ or $({\bf m},{\bf n}) =({\bf n'},{\bf m'})$, and interactions are treated as a perturbation,  scattering processes that change the single-particle modes become off-resonant and atoms remain frozen during the dynamics.  In this case interactions are diagonal in the single-particle basis and for particles in modes $({\bf m},{\bf n})$ they are fully characterized by the interaction energy \begin{eqnarray}
 \label{ueg} &&\hbar U^{\bf nm}_{{\uparrow\downarrow}} =  g \int d^3\vec{x} |\phi_{\bf n}(\vec{x})|^2| \phi_{\bf m}(\vec{x})|^2.
\end{eqnarray}

\textit{Fermions with s-wave interactions in one dimension.}---We begin with the case of  two atoms   tightly confined transversally in their ground state and with dynamics only along the $z-$direction,  where they experience  a 1D harmonic trapping potential with angular trapping frequency $\omega_z$.   The two atoms are initially prepared in the  state $\frac{1}{\sqrt{2}}\left(|n_1,n_2\rangle-|n_2,n_1\rangle\right)|t_{\downarrow\downarrow}\rangle$.

 The atoms are assumed to be in the  Lamb-Dicke regime, with Lamb-Dicke parameter $\eta = k_za_{ho}/\sqrt{2}\ll 1$. $a_{ho}=\sqrt{\hbar/M\omega_z}$ is the harmonic oscillator length, and $k_z$ the projection of the probe laser wavevector along $z$. Mode changes during the laser interrogation can be  suppressed if the laser detuning from the atomic transition, $\delta$,  and the bare Rabi frequency, $\Omega$, satisfy $\delta,\eta \Omega\ll \omega_z$. In this regime the mode-dependence of the Rabi frequencies is $\Omega_n = \Omega e^{-\frac{\eta^2}{2}}L^0_n\left(\eta^2\right)$ \cite{Campbell2009}.
The  Hamiltonian  in the rotating frame of the laser   \cite{Gibble2009, Rey2009, Yu2010, Hazlett2013} under the spin model approximation can be written as $\hat{H}_{sm}^{n_1,n_2}=\hat{H}_{L}^{n_1,n_2}+\hat{H}_{D}^{n_1,n_2}$, where
\begin{eqnarray}
&&\hat{H}^{n_1,n_2}_{L}=\hbar\Delta\Omega^{n_1,n_2}\frac{(\hat{\sigma}^x_1 - \hat{\sigma}^x_2)}{2}-\hbar\bar{\Omega}^{n_1,n_2}{\hat s}_x, \label{smod}\\ \nonumber
&&\hat{H}^{n_1,n_2}_{D} = 2\hbar u_{\uparrow\downarrow}^{n_1,n_2}\hat{P}_{s}-\hbar\delta {\hat s}_z.
\end{eqnarray} $\hat{H}^{n_1,n_2}_{L}$ acts only during the two laser pulses, and $\hat{H}^{n_1,n_2}_{D}$ acts only during the dark time. Here ${\hat s}_{x,y,z}=(\hat{\sigma}^{x,y,z}_1+\hat{\sigma}^{x,y,z}_2)/2$ are collective spin operators and $\bar{\Omega}^{n_1,n_2} = (\Omega_{n_1} + \Omega_{n_2})/2$ is the mean Rabi frequency.  $\Delta\Omega^{n_1,n_2} = (\Omega_{n_1} - \Omega_{n_2})/2$ arises from the excitation inhomogeneity  and can transfer some of the initial triplet population to the singlet, allowing interactions. The interaction energy $\hbar u_{\uparrow\downarrow}^{n_1,n_2}=\hbar U^{\bf nm}_{{\uparrow\downarrow}}$ in Eq.~(\ref{ueg}) with ${\bf n}=\{0,0,n_1\}$ and ${\bf m}=\{0,0,n_2\}$.  We can ignore the detuning and interactions during the laser pulses if the pulses are short compared to the timescales set by those energies. We also ignore single-particle energies which are constants and do not contribute to the dynamics.

The spin model assumptions break down in a harmonic trap due to the degeneracy of the non-interacting two-atom spectrum: even weak  interactions can transfer  atoms   initially in modes $\{n_1, n_2\}$ to the  various degenerate configurations $\{n_1+k, n_2-k\}$ (for integer $k$) during the dynamics. To account for these mode changes, we take advantage of the exact eigenfunctions and eigenvalues of $\hat{H}_{\rm{D}}(\vec{x}_1,\vec{x}_2)$ in Eq.~(\ref{originalH}) for two atoms with $s$-wave interactions in a harmonic trap \cite{Busch1998}. These solutions exploit the separability of the Hamiltonian in the center-of-mass coordinate $R$ and relative coordinate $r$. There is no degeneracy in the relative coordinate degree of freedom. See \cite{SOM} for straightforward expressions for the change of basis. Equivalent expressions are given in \cite{Quemener2011}.


\textit{Ramsey dynamics in the spin model approximation.}---Denoting $\tau$ the Ramsey dark time,  the population difference between the two spin states  measured after the second
pulse takes the generic form
\begin{eqnarray} \label{szdyn}
\langle \hat{s}_z \rangle \left(\tau\right) &=& A(\tau)\cos(\delta \tau) +B(\tau)\sin(\delta \tau) + C(\tau).
\end{eqnarray} 
$A(\tau), B(\tau)$, and $C(
\tau)$ have the form $ A(\tau) = I_1(\tau)f_1+f_2,
B(\tau) = I_2(\tau)f_3,
C(\tau) = I_3(\tau)f_4+f_5$, where $I_i(\tau)$ depend on the dark time physics, and $f_i$ are independent of the dark time physics and depend only on the laser pulse quantities $\{\Delta\theta_{j=1,2}^{n_1,n_2},\bar\theta_{j=1,2}^{n_1,n_2}\}$ (see \cite{SOM}). $\Delta\theta_j^{n_1,n_2}=\Delta\Omega^{n_1,n_2} t_j$ and $\bar\theta_j^{n_1,n_2}=\bar\Omega^{n_1,n_2} t_j$, with $t_{1,2}$ the pulse durations. In the spin model approximation, the dark time functions depend simply on interactions:
$ I_1^{{\rm sm}} = I_3^{{\rm sm}}=\cos(u_{\uparrow\downarrow}^{n_1,n_2}\tau),
I_2^{{\rm sm}} = \sin(u_{\uparrow\downarrow}^{n_1,n_2}\tau)$.
\begin{figure} 
\includegraphics[scale=.6]{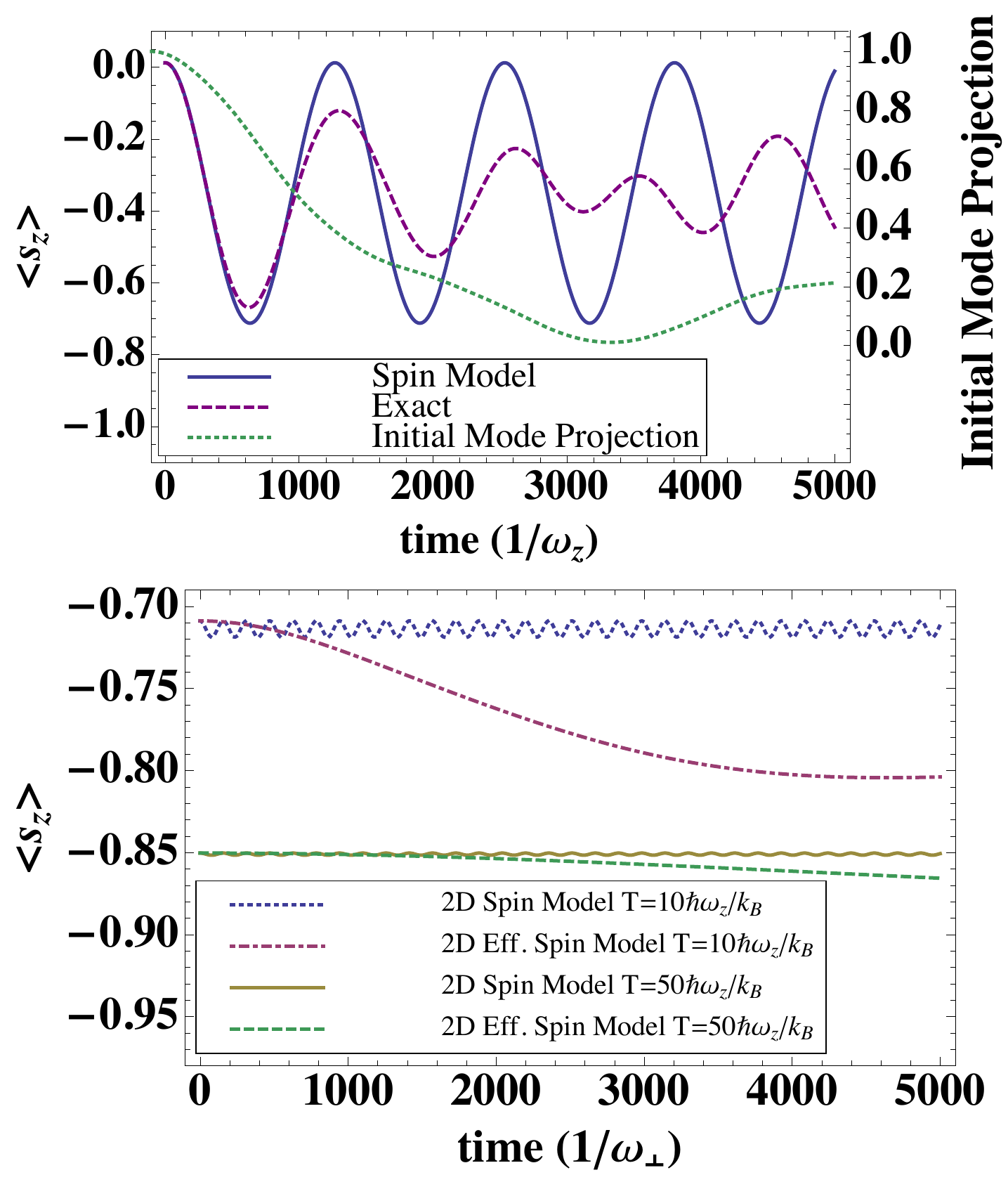}
\caption
{
\label{dynamics}
\raggedright
Ramsey dynamics [see Eq.~(\ref{szdyn})] with $\delta=0$: (a) 1D spin model, exact solution, and projection of population onto initial mode (here  $n_1=10$ and $n_2=0$), with $u_{\uparrow\downarrow}^{1,0} \approx 0.2\omega_z$. Dephasing of the exact dynamics results from mode changes. (b) Thermal averages in 2D: spin model vs. effective spin model, at different temperatures, with $u_{\uparrow\downarrow}^{1,0} \approx 0.04\omega_\perp$. For both figures: $\theta_1=\theta_2=\pi/3$, with thermally-averaged inhomogeneity $\langle \Delta\Omega\rangle/\langle\Omega\rangle = 0.3$. $\theta_i=\Omega t_i$ are bare pulse areas.
}
\end{figure}
\begin{figure} 
\includegraphics[scale=.7]{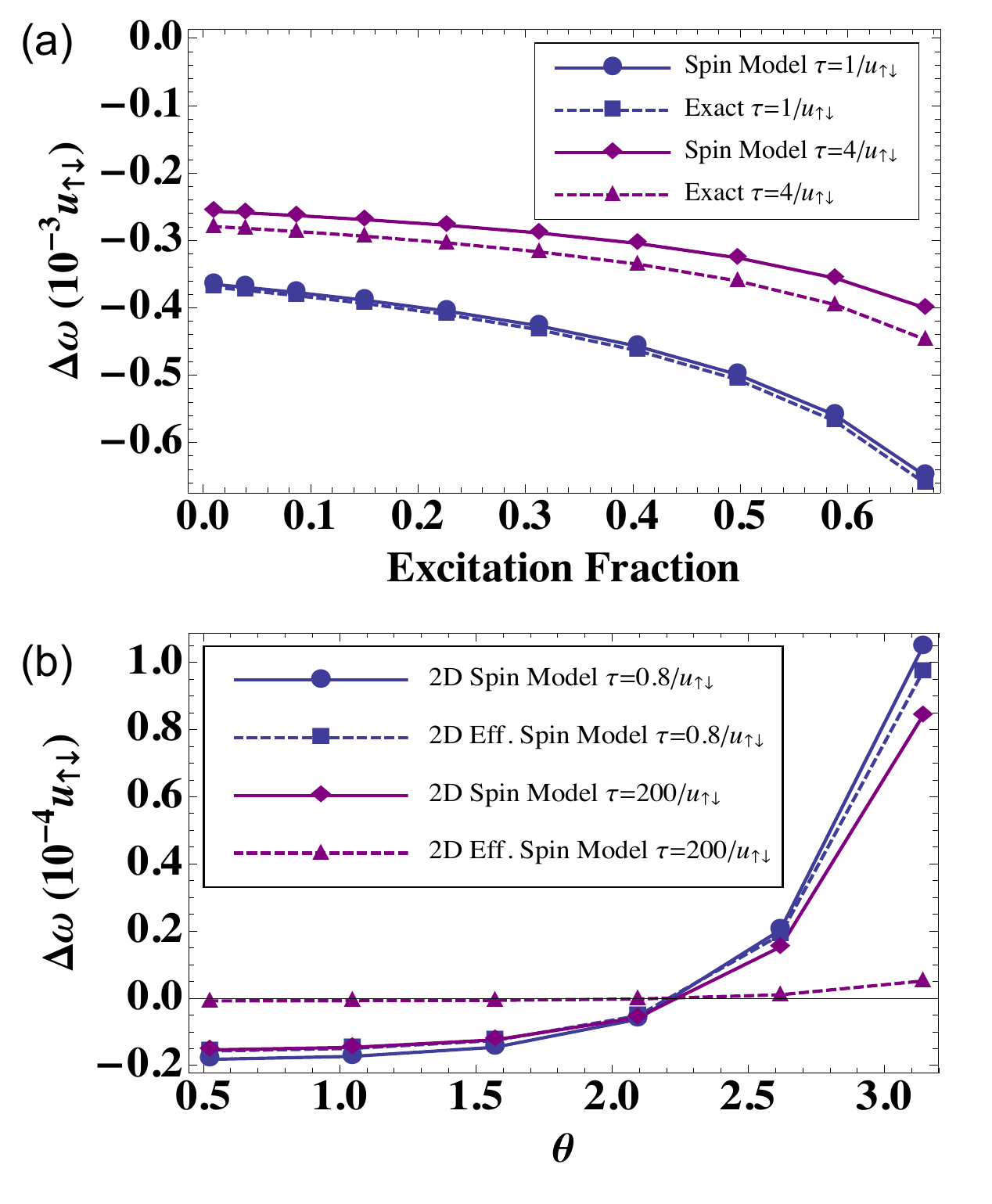}
\caption
{
\label{shifts}
\raggedright
Thermally averaged frequency shifts: (a) 1D spin model and exact solution vs. population excitation fraction (number of atoms in $\uparrow$ divided by the total number of atoms) after the first pulse, at intermediate and long times. Here $u_{\uparrow\downarrow} \equiv u_{\uparrow\downarrow}^{1,0} \approx 0.2\omega_z$. (b) Frequency shifts for 2D spin model vs. effective spin model, with $u_{\uparrow\downarrow} \equiv u_{\uparrow\downarrow}^{1,0} \approx 0.04\omega_\perp$. For both figures: $T =208 \hbar\omega_{\perp}/k_B$,   $\theta=\theta_1=\theta_2$, $\omega_z=\omega_\perp=700\times 2\pi {\rm Hz}$, and thermally-averaged inhomogeneity $\langle \Delta\Omega\rangle/\langle\Omega\rangle = 0.3$. $\theta_i=\Omega t_i$ are bare pulse areas.
}
\end{figure}

\textit{Ramsey dynamics in the weakly interacting regime $(u_{\uparrow\downarrow}^{1,0} \ll \omega_z)$.}---For weakly interacting atoms ($u_{\uparrow\downarrow}^{1,0} \ll \omega_z$), we are able to write the dynamics (beyond the spin model approximation) in a closed analytic form \cite{SOM}. These expressions for the dynamics are exact for times $\tau \ll \omega_z/(u_{\uparrow\downarrow}^{1,0})^2$:
\begin{eqnarray}
I_1^{{\rm exact}} &=& I_3^{{\rm exact}} = 2\displaystyle\sum\limits_{\mathclap{n_r= 0, {\rm even}}}^{n_1+n_2} |d_{n_r}^{n_1,n_2}|^2\cos\left[\frac{\Delta E_s(n_r)}{\hbar}\tau\right], \label{exactabc}\\
I_2^{{\rm exact}} &=& 2\displaystyle\sum\limits_{\mathclap{n_r= 0, {\rm even}}}^{n_1+n_2} |d_{n_r}^{n_1,n_2}|^2\sin\left[\frac{\Delta E_s(n_r)}{\hbar}\tau\right]. \nonumber
\end{eqnarray}
Here $d_{n_r}^{n_{1},n_{2}}$ are the change of basis coefficients defined in \cite{SOM}. Comparing Eq.~(\ref{exactabc}) to the spin model solution, we see the single frequency $u_{\uparrow\downarrow}^{n_1,n_2}$ in the spin-model dynamics gets replaced by a sum over many frequencies $\Delta E_s(n_r)/\hbar$ in the exact dynamics. These frequencies are associated with the first order correction of the eigenenergies  due to interactions \cite{Busch1998}:
$\Delta E_{s}(n_r) = \hbar u_{\uparrow\downarrow}^{1,0} \frac{\Gamma\left(n_r/2+1/2\right)}{\sqrt{\pi}\Gamma\left(n_r/2+1\right)}  \Big(1+
  {\mathcal O}(u_{\uparrow\downarrow}^{1,0}/ \omega_z)\Big)$.
The many frequencies that appear come from the resonant mode-changing processes. States with odd $n_r$ do not experience $s$-wave interactions and do not contribute.

When we compare the exact dynamics to those predicted by the spin model we find that they agree for short times, $\tau u_{\uparrow\downarrow}^{1,0}\ll 1$. The spin model fails at longer times, however, when leakage of population to other modes in the individual-particle coordinate basis becomes significant (See Fig.~\ref{dynamics} (a)). This is reflected in the behavior of the angular frequency shift $\Delta\omega(\tau) $ -- an important quantity for atomic clock experiments -- defined as $
\Delta\omega(\tau)\tau = -\arctan\left[B(\tau)/A(\tau)\right]$, which is the observed change in the atomic transition due to interactions [see Fig.~\ref{shifts} (a)].
 The failure of the spin model at times longer than the inverse interaction strength limits its applicability to model the new generation of  atomic clocks that use ultra coherent lasers \cite{Nicholson2012,Bloom2013}, allowing interrogation  times exceeding a few seconds. A spin model treatment will be insufficient when conditions are such that the atoms see an almost purely harmonic potential.

\textit{Ramsey dynamics in the strongly-interacting regime $(u_{\uparrow\downarrow}^{1,0} \gtrsim \omega_z)$.}---The  spin model  fails when $u_{\uparrow\downarrow}^{1,0} \gtrsim \omega_z$. To maintain the separation between interaction-induced effects and laser-induced effects, we imagine interactions set to be weak during the laser pulses and suddenly increased after the first pulse using for example a Feshbach resonance \cite{Makotyn2013,Hazlett2013, O'Hara2013} 
\footnote{During the pulses we require $u^{1,0}_{\uparrow\downarrow} \ll \Omega$ to ignore interactions, $\eta\Omega \ll \omega_z$ to ignore laser induced mode changes, and $\eta^2\Omega \gg u^{1,0}_{\uparrow\downarrow}$ to populate all of the interacting modes in the singlet. These three conditions cannot simultaneously be satisfied unless $u^{1,0}_{\uparrow\downarrow} \ll \omega_z$.}. For this situation, we can solve for the dynamics, given an initial pair of modes, although there is no closed form solution (the dark time functions $I_i(\tau)$ are more complicated, but the laser dependence through $f_j$ remains the same as in the previous cases). We find that, in the limit of strong interactions ($u_{\uparrow\downarrow}^{1,0} \gg\omega_z$), the population imbalance exhibits periodic oscillations at the axial trapping frequency $\omega_z$, in contrast with the spin model prediction of much faster oscillations at the interaction frequency (see \cite{SOM}). The frequency shift (proportional to this oscillation frequency), saturates to a value on the order of $\omega_z$, instead of increasing without bound. These results reflect the fact that for strong interactions (unitarity), the fermions maximally-repel each other, and the trap energy becomes the only relevant energy scale in the system. This behavior, expected to be a universal result, should apply even in the many-body case as seen in Refs.~\cite{Zwierlein2003,Gupta2003}.

Refs.~\cite{Gibble2009, Hazlett2013} showed that $s$-wave frequency shifts can be cancelled by setting the second pulse area to $\bar{\theta}^{n_1,n_2}_2=\pi/2$. This result, obtained using the spin model, survives the inclusion of resonant mode-changes even for strong interactions during the dark time, since the dependence of the dynamics on the functions $f_i$, and thus $\bar\theta_2^{n_1,n_2}$, remains the same even when interactions are strong.

\textit{Fermions with s-wave interactions in two dimensions.}---For an anisotropic 2D harmonic potential with no accidental degeneracies, the treatment will be similar to the 1D case. An isotropic 2D harmonic potential, however, is more difficult to treat, due to the large degeneracy. In 2D the spin model remains the same as Eq.~(\ref{smod}), with populated modes now ${\bf n_i}=\{n_{xi},n_{yi},0\}$, and interaction energy $\hbar u_{\uparrow\downarrow}^{{\vec n}_1,{\vec n}_2}=\hbar U^{\bf n_1n_2}_{{\uparrow\downarrow}}$ in Eq.~(\ref{ueg}).
To go  beyond the spin model we use polar relative coordinates to eliminate much of the degeneracy. For non-interacting particles, the  eigenfunctions  can be parameterized by quantum numbers $n$ and $m$, with energy $E = \hbar\omega_\perp(2n+|m|+1)$ and angular momentum component $L_z=\hbar m$, where $\omega_{\perp}$ is the 2D oscillator frequency. $S$-wave interactions only affect states with $m=0$, and this subset of states contains no degeneracy (other than the center-of-mass degeneracy). To first order in perturbation theory the interaction energy shift is independent of the radial quantum number $n$: $\Delta E = \frac{g}{2\sqrt{2}\pi^{3/2}a_z a_{\perp}^2}\Big(1+
  {\mathcal O}(u_{\uparrow\downarrow}^{1,0}/ \omega_\perp)\Big)$, where $a_z$ and $a_{\perp}$ are the oscillator lengths along the tightly-confined $z$-direction and the weakly-confined $x$ and $y$-directions, respectively, and in 2D $u_{\uparrow\downarrow}^{1,0} \equiv u_{\uparrow\downarrow}^{{\vec n}_1,{\vec n}_2}$ with ${\vec n}_1 = (1,0,0), {\vec n}_2= (0,0,0)$. This result is striking: despite the large degeneracy in 2D, each interacting state with $m=0$ receives the same energy shift to first order in perturbation theory, and accumulates the same phase  during the dark time. An effective spin model, with diagonal matrix element $2\hbar u_{\uparrow\downarrow}^{{\vec n}_1,{\vec n}_2} $ replaced by $\Delta E$, will be {\it exact} for the $m=0$ states, to first order in the interaction strength. We can replace $\hat{H}^{n_1,n_2}_{D}$ in Eq.~(\ref{smod}) with:
\begin{eqnarray}
\hat{H}^{{\vec n}_1,{\vec n}_2}_{{\rm D, esm}} =\Delta E\hat{P}_{m=0}-\hbar\delta\hat{s}_z,
\end{eqnarray} where $\hat{P}_{m=0}$ projects onto interacting states with $m=0$. For a properly symmetrized initial state $\Psi_{{\bf n_1},{\bf n_2}}(\vec{x}_1,\vec{x}_2)$ in modes $({\bf n_1},{\bf n_2})$, we denote the fraction of the population with $m=0$ in the relative coordinate by ${\mathcal P}^{m=0}_{{\bf n_1},{\bf n_2}}$,which can be calculated as ${\mathcal P}^{m=0}_{{\bf n_1},{\bf n_2}}=\int d^3\vec{x}_1 d^3\vec{x}_2|\Psi_{{\bf n_1},{\bf n_2}}(\vec{x}_1,\vec{x}_2)|^2\delta(\vec{x}_1-\vec{x}_2)=\frac{4\sqrt{2}\pi^{3/2}}{g} \hbar u_{\uparrow\downarrow}^{{\vec n}_1,{\vec n}_2}$, where $u_{\uparrow\downarrow}^{{\vec n}_1,{\vec n}_2}$ is the 2D interaction energy calculated from Eq.~(\ref{ueg}). The dark time dynamics of this effective spin model are simple: $ I_1^{{\rm esm}} = I_3^{{\rm esm}}=(1-{\mathcal P^{m=0}}_{{\bf n_1},{\bf n_2}}) + {\mathcal P}^{m=0}_{{\bf n_1},{\bf n_2}}\cos(\Delta E\tau/\hbar),
I_2^{{\rm esm}} =P^{m=0}_{{\bf n_1},{\bf n_2}} \sin(\Delta E\tau/\hbar)$.

In the original spin model, $\hbar u_{\uparrow\downarrow}^{{\vec n}_1,{\vec n}_2}$ is used as the interaction energy. We see that this parameter appears in the effective spin model to quantify the population of interacting modes (${\mathcal P}^{m=0}_{{\bf n_1},{\bf n_2}}$), instead of their energy. This dramatic result is seen in Fig.~\ref{dynamics}~(b), comparing thermal averages of the previously-implemented spin model with the new effective spin model. Oscillations during the dynamics remain at the same frequency $\Delta E$ at higher temperatures, but the amplitude of the oscillations, proportional to ${\mathcal P}^{m=0}_{{\bf n_1},{\bf n_2}}$, decreases. The previously-implemented spin model, on the other hand, predicts smaller interaction energies (slower oscillations) at higher temperatures. The  frequency shift predicted by the original spin model is only valid  at short times (see Fig.~\ref{shifts}~(b)).

\textit{Summary and Outlook.}---We test the validity of a spin model treatment for Ramsey spectroscopy with exact calculations for two pseudospin-1/2 fermions in a harmonic trap. In 1D the spin model treatment breaks down for dark times on the order of the inverse interaction strength, and for strong interactions. In 2D we find an effective spin model which is exact to first order in perturbation theory, and whose dynamics can be quite different from those predicted by a spin model treatment. Future theoretical treatments of interacting systems probed by Ramsey spectroscopy must take these effects into account to correctly describe dynamics outside of the short-time and weakly-interacting regimes.

\textit{Acknowledgements.}---We thank M. Foss-Feig, G. Campbell, K. Hazzard, A. Ludlow, J. von Stecher, K. O'Hara, M. Martin, and the JILA Sr clock experimental team for feedback. This work was supported by NSF-PIF, NSF JILA-PFC-1125844, NSF JQI-PFC-0822671, NSF IQIM-PFC-1125565, ARO-DARPA-OLE, AFOSR, NIST, and the Lee A. DuBridge and Gordon and Betty Moore Foundations. AK was supported by the Department of Defense through the NDSEG program.

\bibliography{clocks}

\newpage

\begin{widetext}
\section{Supplementary Materials}

\subsection{Change of Basis Formulas}
The exact solutions rely on a change of basis between the individual-particle coordinate basis with wavefunctions $\psi_{n_{1}}(z_1)\psi_{n_{2}}(z_2)$ and the center of mass-relative coordinate basis with wavefunctions $\Psi_{n_R}(R)\Psi_{n_r}(r)$. To convert between these bases we introduce raising operators acting on the vacuum state $|0,0\rangle$, which is the same in both bases: $|n_1=0,n_2=0\rangle = |n_R=0,n_r=0\rangle$. We use the usual form of a raising operator $\hat{a}^\dagger = (\hat{x}-i\hat{p})/\sqrt{2}$ to define
\begin{eqnarray}
\hat{a}^\dagger_R=\frac{1}{\sqrt{2}}(\hat{a}^\dagger_{z_1}+\hat{a}^\dagger_{z_2}), \quad \nonumber
\hat{a}^\dagger_r=\frac{1}{\sqrt{2}}(\hat{a}^\dagger_{z_1}-\hat{a}^\dagger_{z_2})\\ \nonumber
\end{eqnarray}
We can create a particular state out of the vacuum to convert between the two bases:
\begin{eqnarray}
|n_R,n_r\rangle = \frac{(\hat{a}^\dagger_R)^{n_R}(\hat{a}^\dagger_r)^{n_r}}{\sqrt{n_R!n_r!}}|0,0\rangle, \quad \nonumber
|n_1,n_2\rangle = \frac{(\hat{a}^\dagger_{z_1})^{n_1}(\hat{a}^\dagger_{z_2})^{n_2}}{\sqrt{n_1!n_2!}}|0,0\rangle
\end{eqnarray}
These binomials need to be expanded and re-grouped in the form
\begin{eqnarray}
\nonumber \langle R,r|n_R,n_r\rangle = \displaystyle\sum\limits_{i=0}^{n_R+n_r} c_{i}^{n_{R},n_{r}}\psi_i(x)\psi_{n_{R}+n_{r} - i}(y), \quad
\nonumber \langle x,y|n_1,n_2\rangle = \displaystyle\sum\limits_{i=0}^{n_1+n_2} d_{i}^{n_{1},n_{2}}\Psi_i(R)\Psi_{n_{1}+n_{2} - i}(r)
\end{eqnarray}
Grouping the terms, we find
\begin{eqnarray} \label{ci}
c_i^{n_{R},n_{r}} =\sqrt{\frac{i!(n_R+n_r-i)!}{2^{n_R}2^{n_r}n_R!n_r!}}
\displaystyle\sum\limits_{j=\max[0,n_r-i]}^{\min[n_r,n_R+n_r-i]} (-1)^j\left(
\begin{array}{c}
n_R \\
n_R+n_r-i-j \\
\end{array}
\right)
\left(
\begin{array}{c}
n_r \\
j\\
\end{array}
\right)&& \\ \nonumber
d_i^{n_{1},n_{2}}=\sqrt{\frac{i!(n_x+n_y-i)!}{2^{n_x}2^{n_x}n_x!n_y!}}
\displaystyle\sum\limits_{j=\max[0,n_y-i]}^{\min[n_y,n_x+n_y-i]} (-1)^j\left(
\begin{array}{c}
n_x \\
n_x+n_y-i-j \\
\end{array}
\right)
\left(
\begin{array}{c}
n_y \\
j\\
\end{array}
\right) \nonumber
\end{eqnarray}

\subsection{Dependence of Ramsey Dynamics on Laser Pulses}
Eq.~(\ref{szdyn}) in the text gives the generic form of the Ramsey dynamics in terms of functions $A(\tau), B(\tau)$, and $C(\tau)$. These functions depend on the laser pulses through functions $f_i$, given by:
\begin{eqnarray} \label{fi}
f_1 &=& \sin(\Delta\theta_1^{n_1,n_2})\sin(\Delta\theta_2^{n_1,n_2})\cos(\bar\theta_1^{n_1,n_2})\cos(\bar\theta_2^{n_1,n_2}) \\
f_2 &=& \cos(\Delta\theta_1^{n_1,n_2})\cos(\Delta\theta_2^{n_1,n_2})\sin(\bar\theta_1^{n_1,n_2})\sin(\bar\theta_2^{n_1,n_2}) \nonumber\\
f_3 &=& \cos(\Delta\theta_1^{n_1,n_2})\cos(\bar\theta_2^{n_1,n_2})\sin(\Delta\theta_1^{n_1,n_2})\sin(\Delta\theta_2^{n_1,n_2}) \nonumber\\
f_4 &=& -\sin(\Delta\theta_1^{n_1,n_2})\sin(\Delta\theta_2^{n_1,n_2})\sin(\bar\theta_1^{n_1,n_2})\sin(\bar\theta_2^{n_1,n_2}) \nonumber\\
f_5 &=& -\cos(\Delta\theta_1^{n_1,n_2})\cos(\Delta\theta_2^{n_1,n_2})\cos(\bar\theta_1^{n_1,n_2})\cos(\bar\theta_2^{n_1,n_2}) \nonumber
\end{eqnarray}

\subsection{Calculation of Ramsey Dynamics}

Here we briefly sketch the derivation of Eq.~(\ref{exactabc}) in the main text. We begin with two particles in the state $\frac{1}{\sqrt{2}}\left(|n_1,n_2\rangle-|n_2,n_1\rangle\right)|t_{\downarrow\downarrow}\rangle$, where $n_1$ and $n_2$ are quantum numbers for two different (non-interacting) harmonic oscillator modes. A laser pulse is applied, whose action is characterized by $\hat{H}_{L}$ in Eq.~(\ref{smod}). The first pulse has effective pulse area $\bar\theta_1^{n_1,n_2}$ and inhomogeneity $\Delta\theta_1^{n_1,n_2}$. After the first pulse, the state $\frac{1}{\sqrt{2}}\left(|n_1,n_2\rangle+|n_2,n_1\rangle\right)|s\rangle$ becomes populated.  We expand its spatial wave-function into center-of-mass and relative coordinates. To capture the Ramsey dynamics to first order in interaction strength $g$, it is sufficient to leave the wavefunctions unchanged during the dark time. (When we calculate dynamics for strong interactions, we modify these wavefunctions according to Ref.~\cite{Busch1998}.) During the Ramsey dark time, the even relative coordinate modes acquire energies given by $\Delta E_{s}(n_r) = \hbar u_{\uparrow\downarrow}^{1,0} \frac{\Gamma\left(n_r/2+1/2\right)}{\sqrt{\pi}\Gamma\left(n_r/2+1\right)}  \Big(1+
  {\mathcal O}(u_{\uparrow\downarrow}^{1,0}/\omega_z)\Big)$.

After the dark time, we expand the singlet back into individual-particle coordinates which is the convenient basis to calculate the action of a second laser pulse, with effective pulse area $\bar\theta_2^{n_1,n_2}$ and inhomogeneity $\Delta\theta_2^{n_1,n_2}$. The observable $\langle \hat{s}_z \rangle$ is calculated as the population difference between the $|t_{\uparrow\uparrow}\rangle$ and $|t_{\downarrow\downarrow}\rangle$ spin states, summed over each spatial mode. However, only the triplet contributes to the $\langle \hat{s}_z \rangle$ dynamics, and the triplet only contains the original spatial modes $|n_1,n_2\rangle$ and $|n_2,n_1\rangle$, so a major simplification can be made by only summing over these two modes. This simplification is what allows us to calculate the analytic form of Eq.~(\ref{exactabc}).

We can also now understand better why the dependence of the dynamics on the second pulse area is not affected by dark-time interactions. Interactions induce mode changes and introduce new frequencies only to the singlet. The triplet is what determines the dynamics, however, where only the original modes $|n_1,n_2\rangle$ and $|n_2,n_1\rangle$ are present. The second pulse area affects these modes in exactly the same manner as in the spin model treatment of Refs.~\cite{Gibble2009, Rey2009, Yu2010, Hazlett2013}.

\subsection{Strongly Interacting Dynamics}
Fig.~(\ref{szstrong}) shows the Ramsey dynamics predicted by both the spin model and exact calculation for the case of strong interactions. The spin model predicts oscillations at the interaction frequency, which are much faster than the true dynamics which oscillate at the trapping frequency. For fermions, when interactions become very large, the trapping frequency is the only remaining energy scale in the system.
 \begin{figure}
\includegraphics[scale=.5]{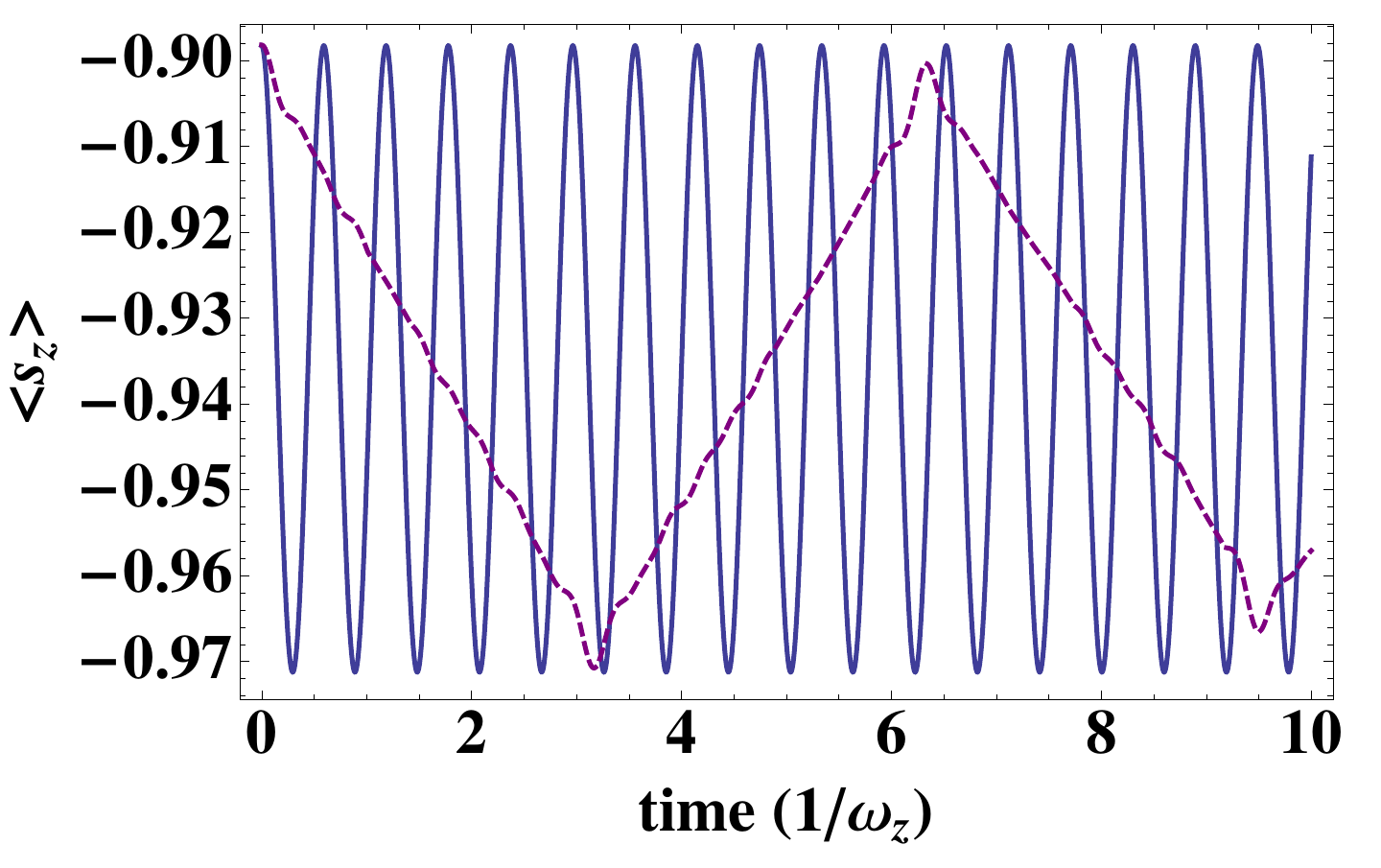}
\caption
{
\label{szstrong}
\raggedright
Ramsey dynamics [see Eq.~(\ref{szdyn})] with $\delta=0$ predicted by the 1D spin model (solid) and the exact solution (dashed) for an  initial ($n_1=6$, $n_2=3$) mode configuration. Strong interactions ($u^{1,0}_{\uparrow\downarrow} = 100\omega_z$) are assumed during the dark time.
}
\end{figure}
\end{widetext}
\end{document}